\documentclass[preprint,prm]{revtex4-2}
\usepackage{geometry}           
\usepackage{graphicx}
\usepackage{subfig}
\usepackage{amsmath,amssymb}
\usepackage{braket}
\usepackage{subcaption}
\graphicspath{{./}}
\usepackage{color}
\usepackage[normalem]{ulem}
\newcommand{\red}[1]{\textcolor{black}{{#1}}}

\newlength{\figwidth}
\newcommand{\fref}[1]{Fig.\,\ref{#1}}
\newcommand{\ffullref}[1]{Figure \ref{#1}}

\newcommand{\efullref}[1]{Equation (\ref{#1})}

\newcommand{\cref}[1]{Ref.\,\cite{#1}}

\begin{document}
\title{\bf {\it Ab initio} calculations of low-energy quasiparticle lifetimes in bilayer graphene
\\
} 

\author{Catalin D. Spataru}\email{cdspata@sandia.gov}
\affiliation{Sandia National Laboratories, Livermore, CA 94551, USA}
\author{Fran\c{c}ois L\'{e}onard}
\affiliation{Sandia National Laboratories, Livermore, CA 94551, USA}


\begin{abstract}
Motivated by recent experimental results  
we calculate from first-principles the lifetime of low-energy quasiparticles in bilayer graphene (BLG). We take into account the scattering rate arising from electron-electron interactions within the $GW$ approximation for the electron self-energy and consider several p-type doping levels ranging from $0$ to $\rho \approx 2.4\times 10^{12}$ holes/cm$^2$. In the undoped case we find that the average inverse lifetime scales linearly with energy away from the charge neutrality point, with values in good agreement with experiments. The decay rate is approximately three times larger than in monolayer graphene, a consequence of the enhanced screening in BLG.
In the doped case, the dependence of the inverse lifetime on quasiparticle energy acquires a non-linear component due to the opening of an additional decay channel mediated by acoustic plasmons.
\end{abstract}

\maketitle

\newpage

Bilayer graphene (BLG) has attracted much attention due to its unique fundamental properties as well as its potential for applications in electronics \cite{Ghazaryan}, chemical sensing \cite{SEEKAEW2017357}, and optoelectronics \cite{photonics9110867}.
For these applications the lifetime of quasiparticles (QPs) plays an important role because it affects the transport properties of electrons through the material and may even determine the current through BLG-based devices. There are several fundamental QP relaxation mechanisms that may contribute to the QP lifetime, such as electron-phonon scattering, impurity scattering and electron-electron scattering. A unique insight into these relaxation mechanisms can be gained via energy-resolved measurements such as angle-resolved/time-resolved photoemission spectroscopy \cite{Bostwick,Moos} or scanning tunneling microspcopy/spectroscopy \cite{Andrei}. Another approach has been demonstrated based on momentum-conserving tunneling spectroscopy 
and recently applied to the study of QP lifetimes in BLG \cite{Tutuc21}, with electron tunneling taking place between two rotationally aligned BLGs separated by a transition-metal dichalcogenide layer. As opposed to a typical device setup with bulk electrodes where the tunneling rate 
 depends directly on the electron density of states 
 in the electrodes, in \cref{Tutuc21} the tunneling rate is dictated by the QP spectral function of the BLGs \cite{MacDonald,Tutuc21}, in particular by the corresponding QP energy-broadening $\Gamma$. These tunneling spectroscopy measurements reveal several interesting aspects of the decay rate $\Gamma$ of QPs with energies $E$ within $\approx \pm200$ meV from the Fermi level ($E_F$): i) near the Fermi level, $\Gamma(E\approx E_F)$ is disorder limited (deduced from its weak temperature dependence), ii) away from $E_F$, electron-electron and/or electron-phonon interactions lead to a linear dependence $\Gamma(E)-\Gamma(E_F) \approx const\times |E-E_F|$, iii) the linear dependence (i.e. the proportionality factor $const$) is insensitive to the doping levels. 

Motivated by the experiments in \cref{Tutuc21}, the goal of this work is to understand the impact of electron-electron interactions on the lifetime of low-energy ($|E-E_F| < 200$ meV) QPs in Bernal BLG, using  {\it ab initio} calculations. Previous {\it ab initio} studies \cite{Park_APL,Park_PRL,Park_NL,Park_PRL_el-ph, Spataru_PRL01} of the QP lifetime in BLG and other graphene-related systems were limited to a higher energy range and did not address the low-energy range relevant to transport.



\begin{figure}[h!]
\subfloat[]
{\includegraphics[width=0.8\textwidth]{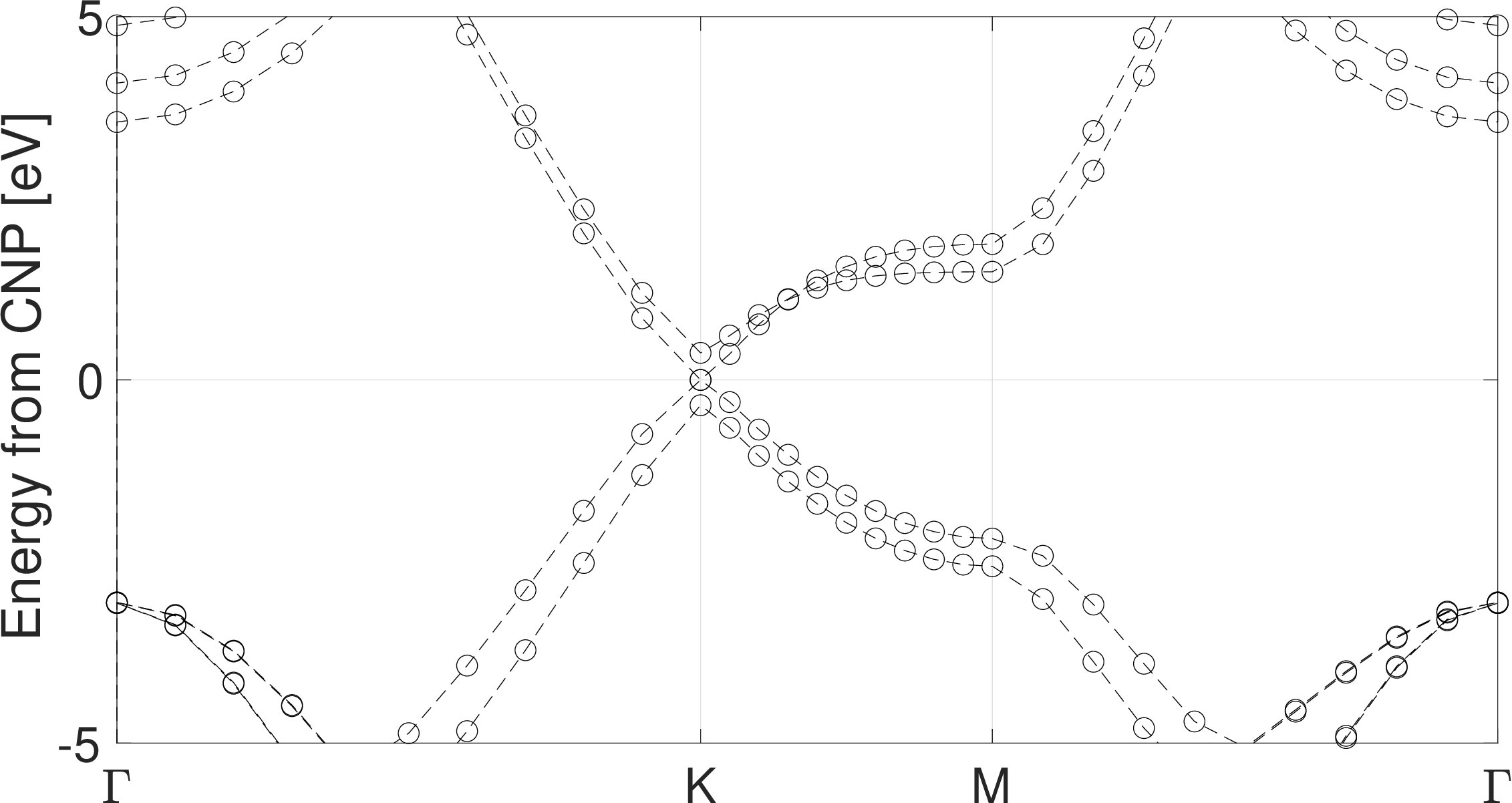}}
\\
\subfloat[]
{\includegraphics[width=0.8\textwidth]{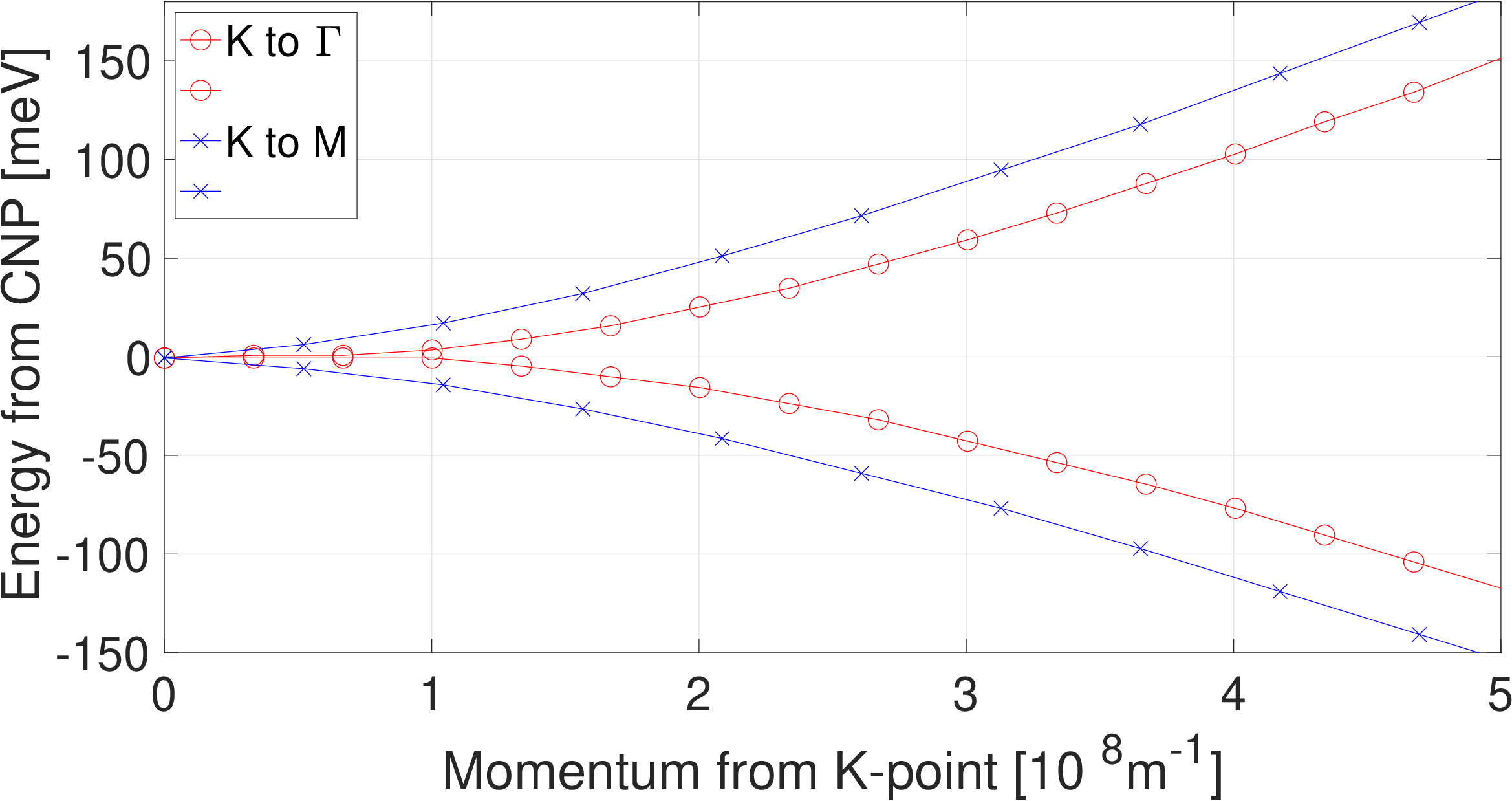}}
\caption{a)  Electronic bandstructure of undoped BLG along high-symmetry $\Gamma-K-M$ directions of the Brillouin zone. b) Low-energy electronic levels along $K \rightarrow \Gamma$ and $K \rightarrow M$ directions. CNP is the charge neutrality point.}
\label{fig:bands}
\end{figure}

\ffullref{fig:bands}(a) shows the electronic bandstructure of BLG calculated within Density Functional Theory (DFT) in the local density approximation (LDA) \cite{Kohn,Ceperley,Perdew} along several high-symmetry directions in the Brillouin zone. The focus of this work will be on low-energy quasiparticles
near the charge neutrality point (CNP) as shown in \fref{fig:bands}(b) for two different directions in the Brillouin zone: $K \rightarrow \Gamma$ and $K \rightarrow M$. The DFT bandstructure is shown for the case of undoped BLG; there are no significant changes that appear in this energy range for the other doping levels considered in this work.


We calculate the QP scattering rate  arising from e-e interactions in BLG by including QP decay processes that involve electron-hole pairs and plasmon excitations.  We calculate the electron self-energy $\Sigma$ within the $GW$ approximation \cite{Louie,SpataruLorin,SpataruGaAs}: $\Sigma=iGW$, and estimate the QP inverse lifetime (FWHM electron linewidth) as: $\tau^{-1}=2 \times Im\Sigma$.

According to the experiment in \cref{Tutuc21}, the temperature dependence of the lifetime of low-energy QPs in BLG is very weak, so in this work we focus on the zero temperature case ($T=0$). In this case QPs can decay only by emitting (as opposed to absorbing) an e-h pair or a plasmon excitation \cite{SpataruLorin}. \efullref{SigmaGW-el} shows the expression for the imaginary part of the electron self-energy of an electronic state  $\ket{nk}$ (band index $n$ and wave vector in the first Brillouin zone $k$) with energy $E^{el}$ above the Fermi level $E_F$ (a quasi-electron):
\begin{equation}
 \mathrm{Im}\Sigma^{el} = -\sum_{m,q,G,G'} M^G_{n,m}(k,q)M^{*G'}_{n,m} \ v(q+G) \ \mathrm{Im}\epsilon_{G,G'}^{-1}(q,E^{el}-E_{mk-q})
\label{SigmaGW-el}
\end{equation}
with $ E_F<E_{mk-q}<E^{el}$.
Similarly, for a quasi-hole excited into an electronic state $\ket{nk}$ with energy $E^h<E_F$, one has:
\begin{equation}
 \mathrm{Im}\Sigma^{h} = \sum_{m,q,G,G'} M^G_{n,m}(k,q)M^{*G'}_{n,m} \ v(q+G) \ \mathrm{Im}\epsilon_{G,G'}^{-1}(q,E_{mk-q}-E^h)
\end{equation}
with $ E^h<E_{mk-q}<E_F$. In these equations $G$ and $G'$ are reciprocal lattice vectors, and $M^G_{n,m}(k,q)=\langle mk\rvert e^{i(q+G)r}\rvert nk-q\rangle$. 

The Coulomb potential $v$ is truncated according to \cref{Sohrab} along the direction perpendicular to the layers (z) to avoid artificial interactions between periodic images (the supercell size along z is $20 \AA$ and the distance between graphene layers in BLG is $3.33 \AA$). The electronic wavefunctions as well as their energies are obtained from the LDA bandstructure.
A plane wave basis with an energy cutoff $60$ Ry was used to represent the Bloch states. In order to obtain well-converged QP lifetimes up to $\pm200$ meV, we considered $16$ bands for BLG ($8$ for graphene) and sampled the Brillouin zone (BZ) with a non-regular grid using a dense sampling near the two non-equivalent $K$-points. The k-point sampling near the $K$-points turns out to be important only in a region of size $\sim 100\times$ smaller than the entire BZ and we obtain converged results by sampling this region with a dense k-point mesh with a mini BZ equal to that of a $1000\times 1000$ regular k-grid. We also find that the inter-valley $K \rightarrow K'$ electronic scattering is not important. 
\red{The dielectric function $\epsilon$ is calculated within the random phase approximation and includes contributions from both intra- and inter- band transitions.} Crystalline local-fields effects in the dielectric matrix were included by summing over $G$,$G'$ up to a cutoff of $3$ Ry and the energy levels were broadened with a small imaginary component of $2$ meV.



\begin{figure}[h!]
\centering
\subfloat[]
{\includegraphics[width=0.8\textwidth]{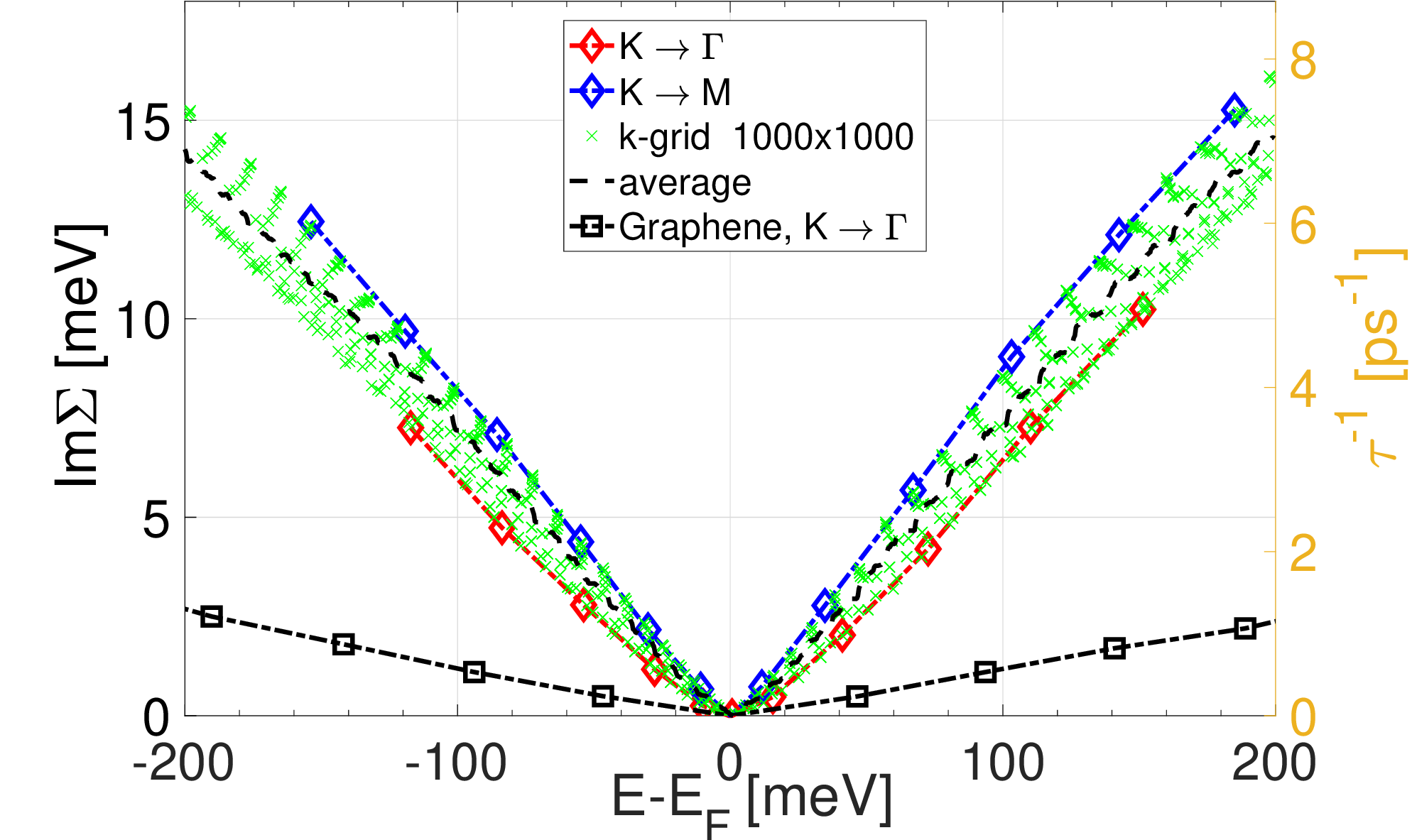}}

\subfloat[]
{\includegraphics[width=0.8\textwidth]{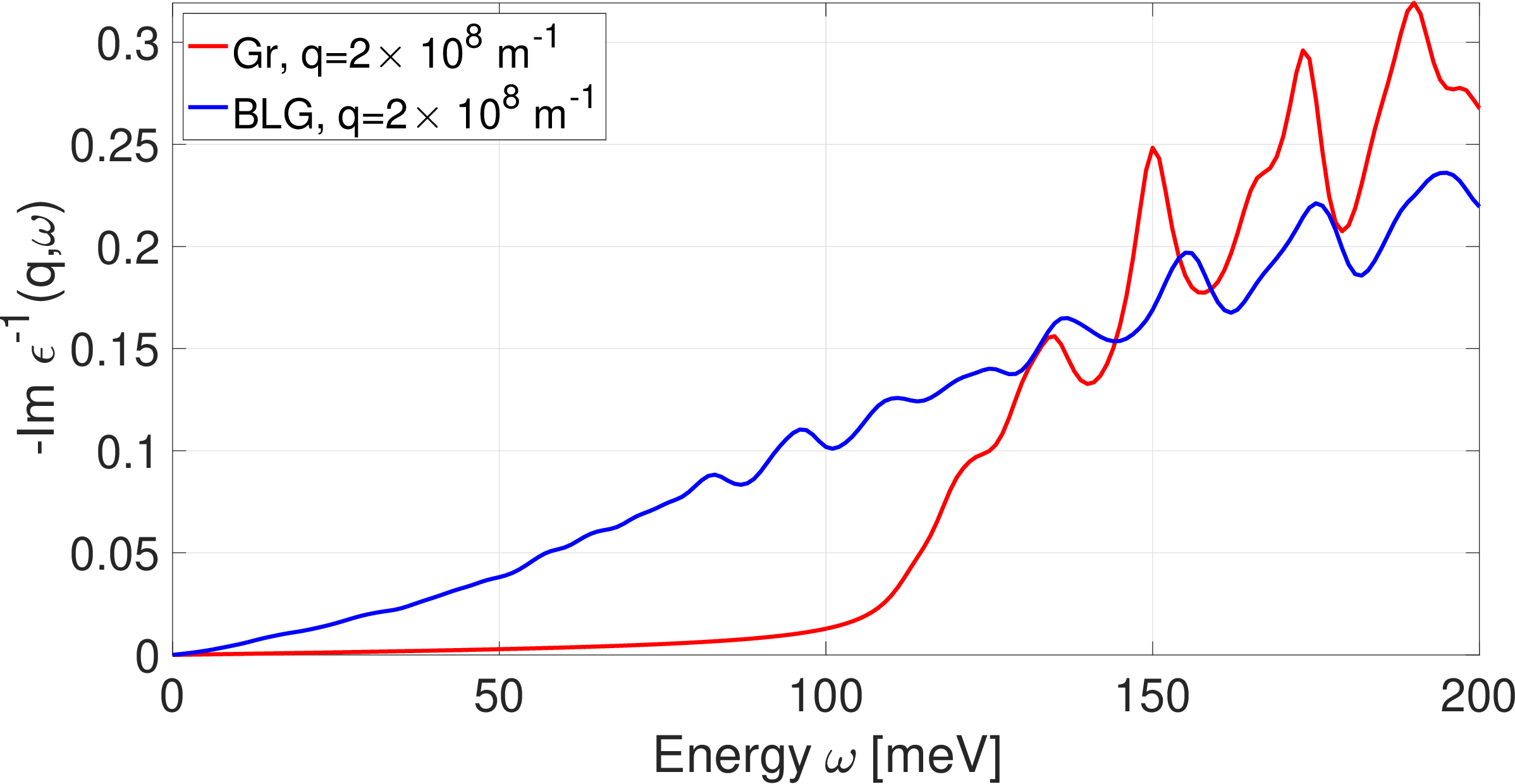}}
\caption{a) Calculated imaginary part of the electron self-energy for undoped BLG and graphene. b) Imaginary part of the inverse dielectric function of graphene and BLG for momentum $q \approx 2\times10^8 ~\text{m}^{-1}$.}
\label{fig:BLG_Graph_und}
\end{figure}

\ffullref{fig:BLG_Graph_und}(a) shows the calculated imaginary part of the electron self-energy $\mathrm{Im}\Sigma$ for undoped BLG, compared to monolayer graphene. 
The lines in red and blue shows $\mathrm{Im}\Sigma$ for BLG calculated for the $K \rightarrow \Gamma$ and $K \rightarrow M$ directions respectively. Because the QP energy dispersion is more pronounced along the $K \rightarrow M$ than along the $K \rightarrow \Gamma$ direction (see \fref{fig:bands}(b)), the QP lifetime is shorter along the $K \rightarrow M$ than along $K \rightarrow \Gamma$ due to the extra phase space corresponding to a QP decay originating along $K \rightarrow M$ and ending along $K \rightarrow \Gamma$.
The green dots also show  $\mathrm{Im}\Sigma$ for BLG calculated for all the QPs near the K-point sampled by a $1000\times1000$ k-point grid, and the black dashed line represents their average over energy intervals of 
width $dE=\pm 10$ meV. We note that the BLG average inverse QP lifetime shows a linear behavior w.r.t. the QP energy away from the $E_F$, except perhaps for a few meV about $E_F$ where the resolution of our calculations does not allow us to draw conclusions and where a quadratic behavior in $E-E_F$ is expected based on Fermi liquid theory.

Our calculations of the lifetimes of low-energy QP in undoped BLG are in very good agreement with the measurements in \cref{Tutuc21}. Indeed, after subtracting from the experimental results the constant disorder contribution $\Gamma(E=E_F)\approx 3$ meV, one obtains $\Gamma^{exp}(E-E_F=125 ~\text{meV})-\Gamma^{exp}(E=E_F)\approx 12$ meV. This compares very well with our calculated results $\mathrm{Im}\Sigma(E=125 ~\text{meV})=9.5 \pm 1$ meV, with the relatively small difference being attributed to other scattering mechanisms not included in our calculations, {\it e.g.} to electron-phonon scattering \cite{Park_APL,Park_PRL_el-ph}. 

For comparison we have also included (black line with square symbols) in \fref{fig:BLG_Graph_und}(a) the calculated $\mathrm{Im}\Sigma$ along the $K \rightarrow \Gamma$ direction for monolayer graphene. We note that the QP lifetime in graphene is about three times longer than that in BLG. This can be explained by the enhanced electronic screening in BLG due to its additional layer. 
Indeed, consider the imaginary part of the inverse dielectric function of BLG and graphene for a given small momentum $q$, as shown in \fref{fig:BLG_Graph_und}(b). One can see that there are more low-energy electron-hole transitions available in BLG than in graphene, a consequence of the fact that the graphene (linear) bands are more dispersive than the BLG (quadratic) bands for energies within $\pm200$ meV from the CNP. In fact, calculating $\mathrm{Im}\Sigma$ for graphene along $K \rightarrow \Gamma$ using the dielectric function of BLG one obtains almost perfect overlap (not shown) with $\mathrm{Im}\Sigma$ for BLG along the same direction. By contrast, in \cref{Park_APL} it was found that in BLG the scattering rate $\tau^{-1}$ arising from electron-electron interactions is smaller than that in graphene by $20$ to $40\%$ on average.  The difference may be explained by the fact that in \cref{Park_APL} the focus was on a QP energy range ($\sim 1 $ eV) much higher than the one in the present work ($\sim 100 $ meV) . 

\begin{figure}[h!]
\setcounter{subfigure}{2}
\centering
\subfloat[]
{\includegraphics[width=0.45\textwidth]{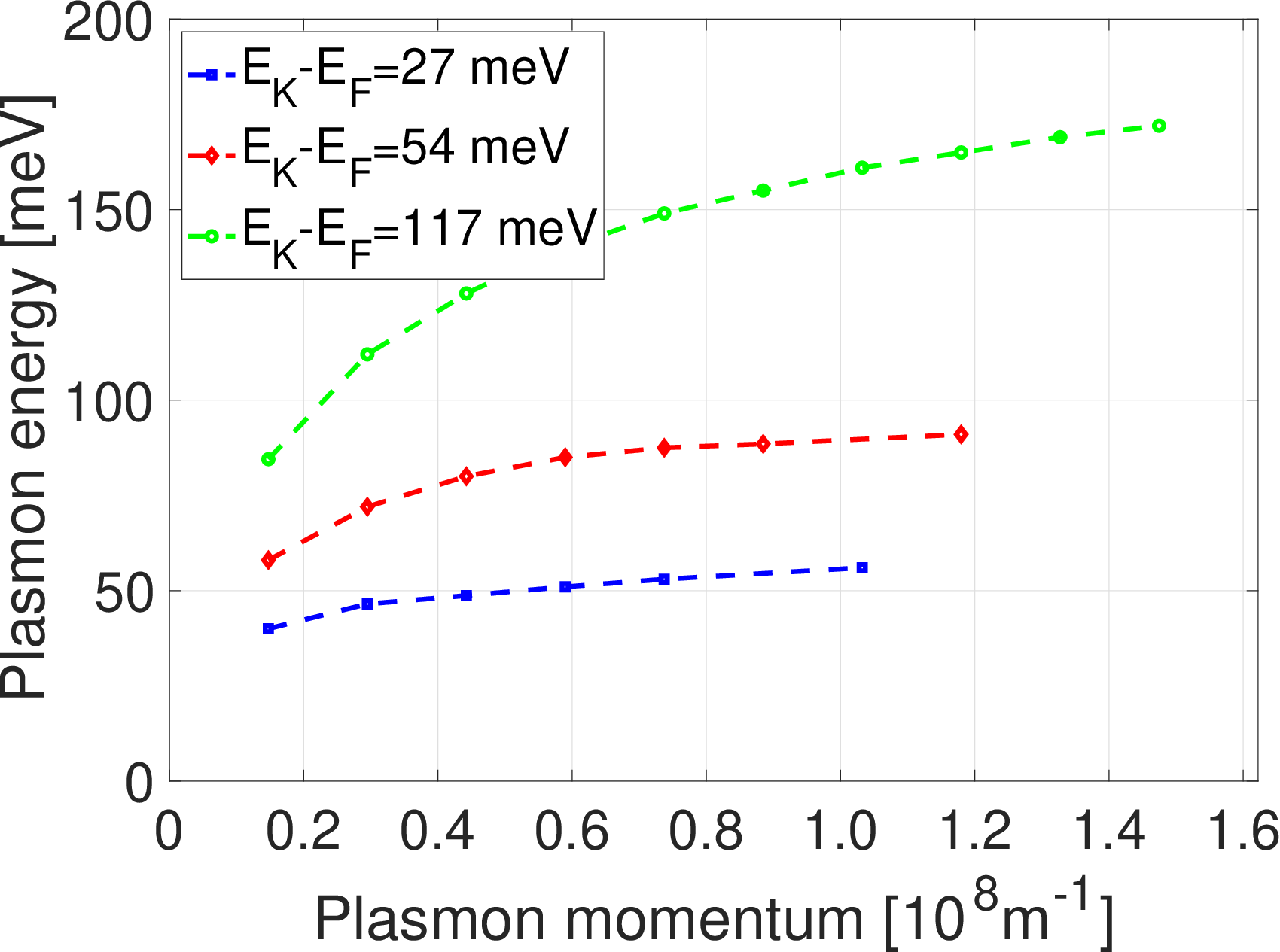}}
\centering
\subfloat[]
{\includegraphics[width=.6\textwidth,trim={9cm 5cm 9cm 5cm},clip]{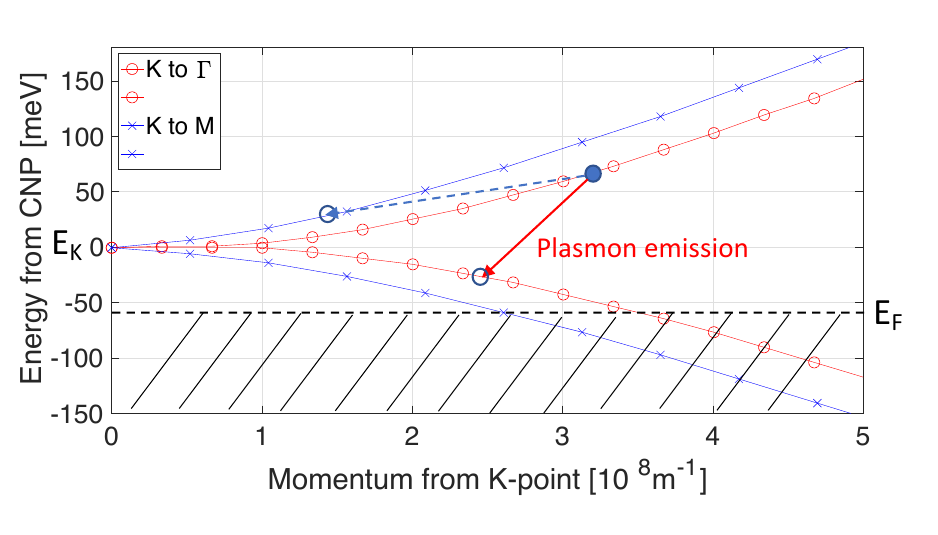}}
\caption{a) Calculated imaginary part of the electron self-energy in BLG for several doping levels. Each line represents an average over all the QP states sampled by the $1000\times 1000$ k-point grid with energies within intervals of width $dE=\pm 10$ meV. \red{The inset shows the calculated Im$\Sigma$ for quasi-electrons with momentum along the $K\rightarrow\Gamma$ direction and doping level $E_K-E_F=54$ meV.} 
b) Imaginary part of the inverse dielectric function $\epsilon^{-1}(q,\omega)$ for several momenta $q$ and doping level $E_K-E_F=117$ meV. c) Plasmon dispersion in doped BLG for several doping levels. Plasmon momentum along the $\Gamma \rightarrow M$ direction.
d) Electronic bands of doped BLG ($E_K-E_F=54$ meV) showing a sketch of a quasi-electron decay process enabled by plasmons.}
\label{fig:avg_Im_All_Dop}
\end{figure}

\ffullref{fig:avg_Im_All_Dop}a) shows the calculated average imaginary part of the electron self-energy for p-type doped BLG for several positions of the Fermi level $E_F$ with respect to the energy of the K-point $E_K$ (the CNP). Assuming a hole effective mass $m^*\approx 0.05 m_e$ \cite{Zou} the highest doping level we considered ($E_F=117$ meV) corresponds to $\rho\equiv E_F m^*/\hbar^2\pi\approx 2.4\times 10^{12}$ holes/cm$^2$. 
While $\mathrm{Im}\Sigma$ shows approximate linear behavior with $|E-E_F|$ in the undoped case ($E_F=E_K$), the inverse lifetime becomes overall smaller in the doped case and acquires a non-linear component that shows up for quasi-electrons with energies approximately twice as large as $E_K-E_F$.

To understand the origin of the aforementioned non-linear component we need to first consider the emergence of acoustic plasmons in doped BLG. Indeed, based on simple dimensionality arguments, free carriers in 2D systems lead to the formation of plasmons with energy approaching zero in the long wavelength limit: $\omega_{pl}(q) \sim \sqrt{q}$. \ffullref{fig:avg_Im_All_Dop}b) shows the imaginary part of the inverse dielectric function $\mathrm{Im}\epsilon^{-1}(q,\omega)$ for several momenta $q$ along the $\Gamma \rightarrow M$ direction and doping level $E_K-E_F=117$ meV. The peak in $\mathrm{Im}\epsilon^{-1}(\omega)$ indicates a plasmonic excitation, whose weight decreases as $q$ increases -due to overlap with interband electron-hole transitions \cite{Chakraborty,dasSarmasBLG}- while its energy saturates towards a value close to $2\times(E_K-E_F)$.
\ffullref{fig:avg_Im_All_Dop}c) shows the energy dispersion of the acoustic plasmons $\omega_{pl}(q)$ with $q$ along the $\Gamma \rightarrow M$ direction. We note that the $q$-range where the plasmons can be identified increases with doping and that $\omega_{pl}(q)$ increases linearly with the density of free carriers. The density of plasmons is highest for plasmon energies close to saturation, {\it i.e.} close to $2\times(E_K-E_F)$.

Next, we focus on a particular doping level such as the one characterized by $E_K-E_F=54$ meV (lines in red in \fref{fig:avg_Im_All_Dop}(a),(c)). To understand the origin of the peak in the QP decay rate for quasi-electrons with energies $\approx 100$ meV above $E_F$, we consider an electronic state in the BLG banstructure - indicated by the filled-circle in \fref{fig:avg_Im_All_Dop}(d). This quasi-electron, occupying a state in the conduction band, can decay into an available empty state from either the conduction band \red{-indicated by the blue arrow in \fref{fig:avg_Im_All_Dop}(d)}, or from the valence band \red{-indicated by the red arrow in \fref{fig:avg_Im_All_Dop}(d)}.
\red{The first process (decay into conduction band) does not contribute to the peak in Im$\Sigma$, as demonstrated in the inset of \fref{fig:avg_Im_All_Dop}(a) which shows Im$\Sigma$ for quasi-electrons with momentum along $K\rightarrow\Gamma$, calculated in two ways: i) full calculation with decay into either conduction or valence bands, and ii) with decay constrained to the conduction band only.}
 The latter process \red{(decay into valence band)} can be facilitated by the emission of a low-momentum 
plasmon as indicated by the arrow in \fref{fig:avg_Im_All_Dop}(d). This implies that an extra decay channel (mediated by plasmons) opens up for quasi-electrons. (An equivalent decay channel opens up for quasi-holes in the case of n-type doping.) A similar correlation between the QP energy where the non-linear features in the decay rate of quasi-electrons appear (see \fref{fig:avg_Im_All_Dop}(a)) and the plasmon energy that corresponds to a high density of plasmons (see \fref{fig:avg_Im_All_Dop}(c)) exists for all the other doping levels.

We note that experimentally the linear behavior of $\tau^{-1}$ with QP energy away from $E_F$ is insensitive to the doping level, {\it i.e.} the non-linear features that we calculate are not observed experimentally \cite{Tutuc21}. This suggests that with increasing doping other scattering channels, besides the electron-electron scattering channel considered here, may become important. 
One may ask if the electron-phonon scattering channel may explain the difference between our calculations and experiment. According to previous {\it ab initio} studies of graphene and BLG \cite{Park_PRL_el-ph,Park_APL}, only optical phonons are effective in electron-phonon scattering. 
Because the typical energy of optical phonons is $\approx 0.2$ eV \cite{phonons_BLG_Gr}, their contribution to the imaginary part of the electron self-energy turns out to be very small ($< 1$ meV) \red{within the adiabatic approximation} at $T=0$ in both undoped and doped ($|E_K-E_F|=1$ eV) cases for quasiparticles with energies within $\approx200$ meV from the Fermi level \cite{Park_PRL_el-ph,Park_APL}. 
Thus, the aforementioned non-linear features in the calculated $\tau^{-1}$ are expected to persist even when including electronic transitions through phonon emission. \red{We note however that the adiabatic approximation is valid when the characteristic phonon frequency is much smaller than the typical energy of the electronic excitations (plasmons). This may not be valid in doped BLG where the plasmon dispersion is acoustic; in particular for the doping levels considered in this work the typical plasmon energy is smaller than the characteristic optical phonon energy ($\approx 200$ meV) in BLG, as seen in \fref{fig:avg_Im_All_Dop}(c). Thus, non-adiabatic effects in the electron-phonon interaction \cite{FGiustino17} may be important and deserve to be included in a future study as a possible mechanism that could improve the agreement between theory end experiment.}

In conclusion, we have presented a first principles study of the lifetime of low-energy QP in BLG with contribution from the electron-electron scattering channel. For undoped BLG we find that $\tau^{-1}$ shows linear dependence with the QP energy $E$ for $|E-E_K|>$ few meV, while the QP lifetime is significantly shorter than in graphene due to enhanced electronic screening. 
For doped BLG, the opening of an additional decay channel mediated by acoustic plasmons leads to a non-linear feature in $\tau^{-1}$.  
While the calculated lifetime is in very good agreement with experiment \cite{Tutuc21} for the undoped case, our study suggests - in combination with the experimental results in \cref{Tutuc21} - that other relaxation mechanisms (not arising purely from electron-electron interactions) may become important as doping increases.
\red{We suggest that non-adiabatic effects in the electron-phonon interaction may be important and should be considered in future work. In addition, the experiments to extract the carrier lifetime rely on a
new approach that must satisfy specific conditions on the transport. Further
theoretical analysis may be needed to fully understand the factors that
may influence the extracted/calculated lifetimes.}

This work was supported by the LDRD program at Sandia National Laboratories (SNL). SNL is a multimission laboratory managed and operated by National Technology and Engineering Solutions of Sandia, LLC., a wholly owned subsidiary of Honeywell International, Inc., for the U.S. Department of Energy's National Nuclear Security Administration under contract DE-NA0003525.
The views expressed in the article do not necessarily represent the views of the U.S. Department of Energy or the United States Government.

\clearpage

\clearpage

\clearpage
\begin{figure}[h!]
\centering
\subfloat[]
{\includegraphics[width=1.2\textwidth,trim={3cm 1.4cm 1cm 1cm},clip]{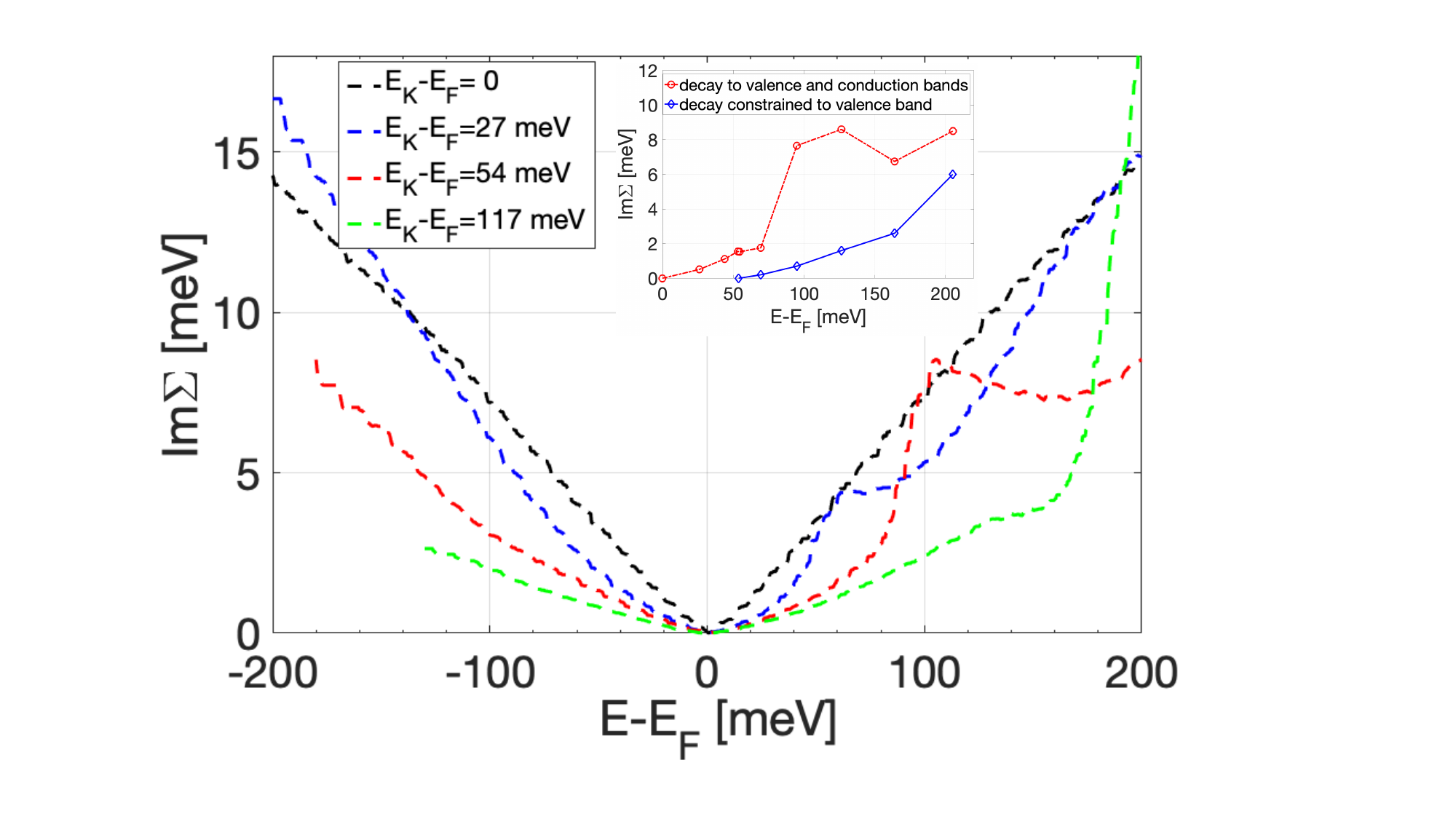}}

\centering
\subfloat[]
{\includegraphics[width=0.9\textwidth]{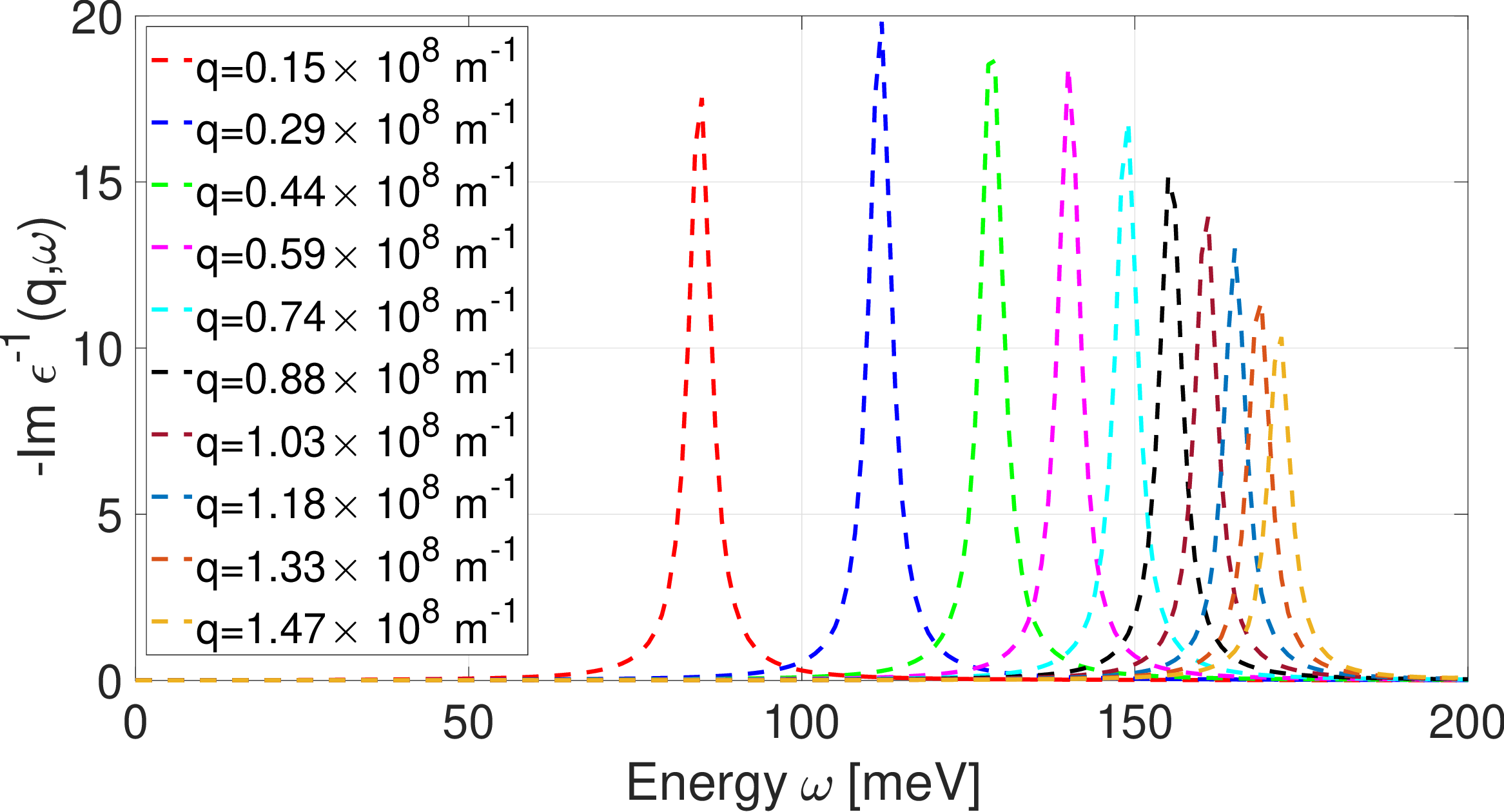}}
\end{figure}

\end{document}